# Factors influencing thermal radiative properties of metals

Kralik T., Hanzelka P., Musilova V., Srnka A.
Institute of Scientific Instruments of the ASCR, v.v.i., Kralovopolska 147, Brno, 612 64 Czech Republic

## ABSTRACT

Data on thermal absorptivity and emissivity are important for calculations of heat flows in cryogenic devices. Although thermal radiative properties of materials used in cryogenics are measured, published and used for more than sixty years we have noticed that sometimes users of that data don't well understand the significance of individual factors influencing thermal radiative properties.
The sensitivity of the thermal radiative properties of metals to the condition of the surface and the difficulty of measurement, especially at low temperatures is probably the reason of dispersion of the published values. The effect of the material and its purity, finishing (machining, mechanical or chemical polishing) and of the protective coatings or accidental layers (water, oxidation) on the thermal radiative properties is not always obvious. Influence of the material, surface treatment, roughness and layers on the surface is discussed on the basis of our 20 year experience with measurement of the emissivity and absorptivity in the range from 20 K to 300 K.

## 1. INTRODUCTION

The design of cryogenic apparatuses for industry, medicine, laboratory experiments or space exploration is always a compromise between the demands on cryogenic performance and limitations of available technologies. To be able to recognize what concession could be done the influence of used technologies and treatments on material properties should be known.
Thermal radiation inside a cryogenic apparatus is a significant mechanism of heat transfer with high impact on the overall properties of the apparatus like evaporation rate, cryocooler load or radiative heating of critical parts, e.g. samples or temperature sensors.
Thermal radiation penetrates into, and interacts with very thin surface layer (tens of nanometres) of a metal. The sensitivity of thermal radiative properties to the state of the material near the surface is thus one of the reasons for large dispersion of data published for the same type of material (Musilova et al., 2005, Musilova et al., 2007).
The Group of Cryogenics and Superconductivity at the Institute of Scientific Instruments (ISI) has a 20 year experience with measurement of total hemispherical emissivity and absorptivity of miscellaneous materials at temperatures ~20 K - 300 K. In this paper the effects of various factors influencing the thermal radiative properties of highly reflecting metals are discussed.

## 2. MEASUREMENT METHOD

The device and the method used at ISI for measurement of thermal radiative properties are described in detail in (Kralik, 2004) and here will be only shortly introduced. The used apparatus enables to measure total hemispherical absorptivity and emissivity at temperatures ~20 K - 300 K. The principle is based on measurement of radiative heat flow $Q_R$ between two parallel surfaces with different temperatures, a sample surface and a reference surface with high absorptivity and emissivity. The heat transferred by thermal radiation from the heated surface (radiator at the temperature $T_R$) to the cold surface (absorber at the temperature $T_A$) flows through a thermal resistor (heat flowmeter) into a heat sink (LHe bath with stabilized temperature). The absorbed radiative heat flow is evaluated from the temperature drop $T_A$ - $T_B$ measured on the heat flowmeter. Calibration of the heat flowmeter and measurement of the emissivity or absorptivity are conducted in the same instrument setup.
Relation between $Q_R$ and mutual emissivity $\varepsilon_{RA}$ of the sample and the reference surface is given by eq. 1 (Siegel and Howell, 2002), where $A$ denotes the sample area and $\sigma$ the Stefan-Boltzmann constant. Emissivity $\varepsilon_{RA}$ results from the thermal radiative properties both of the radiator and the absorber (eq. 2) (Siegel and Howell, 2002). Total hemispherical absorptivity $\alpha(T_R, T_A)$ is measured, when the sample is cold and irradiated with the heat emitted by the heated reference surface. On the contrary, the sample is heated and the reference surface is cold during the total hemispherical emissivity $\varepsilon(T_R)$ measurements.

$$\varepsilon_{RA} = \frac{Q_R}{A\sigma(T_R^{\ 4} - T_A^{\ 4})} \quad (1)$$

$$\frac{1}{\varepsilon_{RA}} = \frac{1}{\varepsilon_R} + \frac{1}{\alpha_A} - 1 \quad (2)$$

$$\frac{1}{\varepsilon_R} = \frac{1}{\varepsilon_{RA}} - \frac{1}{\alpha_{REF}} + 1 \qquad \frac{1}{\alpha_A} = \frac{1}{\varepsilon_{RA}} - \frac{1}{\varepsilon_{REF}} + 1 \quad (3)$$

The emissivity $\varepsilon_R$ or absorptivity $\alpha_A$ of the sample is evaluated from eq. 3 by using known emissivity or absorptivity of the reference surface.
The sample is a disk with 40 mm in diameter. The reference sample of the same shape is coated with a layer with high emissivity and absorptivity.

## 3. MATERIALS AND TREATMENTS

In the following tables the materials of the samples, their working or finishing and composition of chemical baths used for polishing, passivation and pickling are specified.

Table 1. Sample materials

| Material | Characterization |
|---|---|
| Cu | Cu 99.5 % as 1 mm thick sheet: Copper produced according to the Czech standard CSN 42 3005. Composition according to the standard: impurities less than 0.5 %; max. contents of individual impurities are Fe 0.05 %, Al 0.05 %, Pb 0.1 %, Sn 0.15 %, As 0.1 % and Sb 0.08 %. |
| Al | Al 99.5 % as 1 mm thick sheet: Aluminium produced according to the Czech standard CSN 42 4005.21, equivalent to EN 573-3 [AW1050A]. Composition according to the standard: impurities less than 0.5 %; max. contents of individual impurities are Cu 0.05 %, Fe 0.4 %, Si 0.3 %, Ti 0.05 % and Zn 0.07 %. |
| Ni | Rolled 1 mm thick Ni sheet, 99.5 %. |

Table 2. Description of the samples. The second number (history) denotes a subsequent treatment of the same sample.

| Sample No./history | Material | Surface material | Characterization | Roughness Ra [µm] |
|---|---|---|---|---|
| 04/3 | Al | Al | Chemically polished (bath 2) and for half a year exposed to air in laboratory | --- |
| 04/4 | Al | Al | Again chemically polished (bath 2) | --- |
| 04/4 | Al | Al | Chemically polished (bath 2) with in situ deposited 85 nm layer of water ice | 0.19 |
| 08/1 | Al | Al | Unfinished, from the shelf, with natural oxide layer | 0.36 |
| 10/1 | Al | Al | Abraded with stainless steel wool | 1.33 |
| 11/1 | Al | Al | Finely turned on a lathe | 0.64 |
| 12/1 | Al | Al | Mechanically polished to high gloss | 0.04 |
| 14/1 | Cu | Cu | Chemically polished (bath 1, 3 minutes) | --- |
| 14/2 | Cu | Cu | Chemically polished (bath 1) and aged (exposed for 6 months to clean air) | --- |
| 14/3 | Cu | Cu | Chemically polished (bath 1, 2 minutes) | 0.27 |
| 15/1 | Cu | Cu | Surface layer of 0.1 mm etched off in concentrated HNO3 | 0.46 |
| 22/1 | Cu | Cu | Machined finely on the lathe | --- |
| 22/2 | Cu | Cu | Sample 22/1 vacuum annealed at 650°C for 30min | 1.29 |
| 27/1 | Cu | Cu | Lapped and mechanically polished to high gloss | 0.03 |

Table 2 continued. Description of the samples. The second number (history) denotes a subsequent treatment of the same sample.

| | | | | |
|---|---|---|---|---|
| 38/1 | Cu | Cu | Chemically polished (bath 1) and passivated (bath 3) at room temperature, 1 minute) | 0.24 |
| 41/1 | Cu | Ni chemically deposited | 10 µm layer of the alloy 90 % Ni and 10 % P, bath Schlötter (Ireland) | 0.21 |
| 42/1 | Cu | Ni galvanically deposited | 10 µm layer of pure Ni, bath Schlötter (Ireland) | 0.24 |
| 46/1 | Cu | Au galvanically deposited | 2 µm layer, contains of about 0.6 % of Co, commercial bath | 0.12 |
| 47/1 | Ni | Ni | Ni sheet pickled (bath 4 at 35°C, 2 minutes) | 0.16 |
| 48/1 | Cu | Cu | Unfinished, exposed to aggressive air in chemical laboratory | 0.24 |

Table 3. Composition of used chemical baths

| Bath No. | Usage | Composition |
|---|---|---|
| 1 | Chem. Cu polishing | $H_3PO_4$ 45 ml, $CH_3COOH$ 35 ml, $HNO_3$ 20 ml. All acids with maximal concentration. Bath at room temperature. |
| 2 | Chem. Al polishing | $H_3PO_4$ 80 ml, $HNO_3$ 12 ml, $H_2SO_4$ 8 ml, $Cu(NO_3)_2$ 0.2 g. All acids with maximal concentration. Bath at 90°C - 100°C for 30 s. Further cleaned with 20 % $HNO_3$, rinsed with water and dried. |
| 3 | Passivation of Cu | $H_2O$ 100 ml, $K_2Cr_2O_7$ 10 g, $H_2SO_4$ 0.5 ml. |
| 4 | Pickling of Ni | $H_2O$ 20 ml, $H_2SO_4$ 30 ml, $HNO_3$ 45 ml, NaCl 0.6 g. |

## 4. FACTORS AFFECTING THE THERMAL RADIATIVE PROPERTIES

The following chapters are devoted to the individual factors affecting the total hemispherical absorptivity of samples described in the Table 2.

### 4.1. Temperature of the material

While total hemispherical emissivity $\varepsilon(T_R)$ is temperature dependent material property, absorptivity $\alpha(T_A, T_R)$ depends on both material temperature $T_A$ and on frequency spectrum of impacting black body radiation (i.e. on the temperature $T_R$ of radiation). Notice, that in accord with Kirchhoff's law, when temperature $T_R$ approaches the value of $T_A$ the difference between emissivity and absorptivity diminishes, $\varepsilon(T_R)= \alpha(T_A, T_R)$, $T_A = T_R$.

In the case of metallic materials at low temperatures, values of emissivity increase with $T_R$ and absorptivity values increase with both $T_R$ and $T_A$ (Siegel & Howell, 2002; Musilova et al., 2005). The reason for such dependence resides in the relation of optical properties to electrical resistivity (Siegel & Howell 2002, Domoto, Boehm & Tien, 1970) due to dominated interaction of radiation with free electrons in metal at respective temperatures of radiation. When optical properties (electrical resistivity of a metal) are temperature independent, emissivity and absorptivity depend only on temperature $T_R$ of radiation and they equal as a consequence of Kirchhoff's law. This is the case of pure metals in the region of dominating constant residual resistivity or anomalous skin effect at lowest temperatures (Fig. 1, $T_R < 60$ K) and of low purity metals and alloys in broad temperature region (Fig. 9 in Musilova et al., 2005, e.g.). For pure metals the emissivity at higher temperatures $T_R$ is higher than the absorptivity of the same sample (Fig. 1). This difference consists in difference between optical properties (resistivity) of pure metal at low temperatures (sample is at low temperature $T_A$ during absorptivity measurement) and high temperature (sample temperature $T_R$ varies from low to high temperature during emissivity measurement).

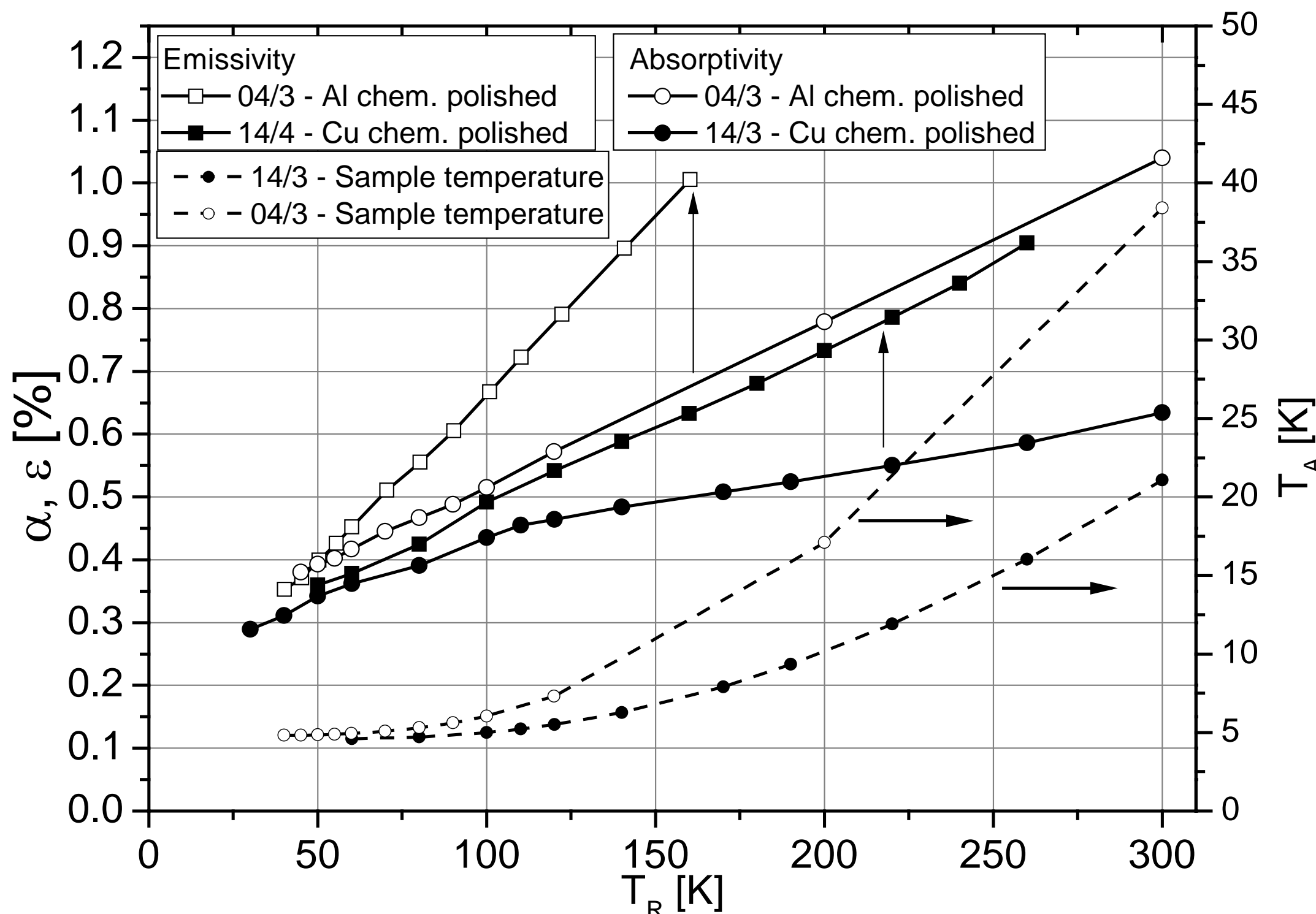

Fig. 1 Emissivity and absorptivity of chemically polished copper and aluminium, both with technical purity (99.5 %). The right hand side axis shows the temperature of the samples $T_A$ during absorptivity measurement. $T_R$ is the temperature of the sample in emissivity measurement and the temperature of the source of radiation in measurement of absorptivity. The observed increase in $T_A$ is given by limitation of used method. Nevertheless, the nearly zero difference between values of absorptivity and emissivity at temperatures $T_R \leq 50$ K shows that resistivity of samples does not depend on temperature $T_A$ below 50 K and thus measured values of absorptivities should not be influenced by increase of temperature $T_A$ up to 50 K.

## 4.2. Material purity

4.2.1. Is the metal of higher purity better?

Thermal radiation (far infrared electromagnetic field) impacting metal surface penetrates into a thin "skin depth" where it interacts predominantly with free electrons in a similar way as voltage applied to a bulk metal. Energy of accelerated electrons then partly dissipates in a form of Joule heat (is absorbed) by collisions with vibrating crystal lattice, impurities and crystallographic defects in the skin depth, and also by collisions with the surface itself. In general, the absorptivity increases with DC metal resistivity. The low temperature DC electrical resistivity is lowest in pure metals, characterized by high value of the RRR. Actually, when for example, the term $\rho_S$ dominates in Eqs. (4) and (5) the emissivity and absorptivity does not follow the decrease of DC resistivity with temperature or with increasing material purity.

In Fig. 2 the calculated absorptivities of copper are compared with measured absorptivity of chemically polished technical copper (99.5 %) characterised with RRR = 40. The measured values of absorptivity agree within ~25 % of their value with the approximate theoretical values. In Fig. 2 the plotted theoretical curves for various values of RRR show that using copper with higher than technical purity, e.g. RRR > 40, is not effective. It is possible to do similar conclusion for chemically polished samples of aluminium. This statement is also supported experimentally by agreement of absorptivities $\alpha$(5 K, 300 K) obtained by us for chemically polished Cu and Al, both of 99.5 % purity, with the results obtained by Ramanathan (Ramanathan, 1952) for electro-polished samples of Cu and Al of much higher purity.

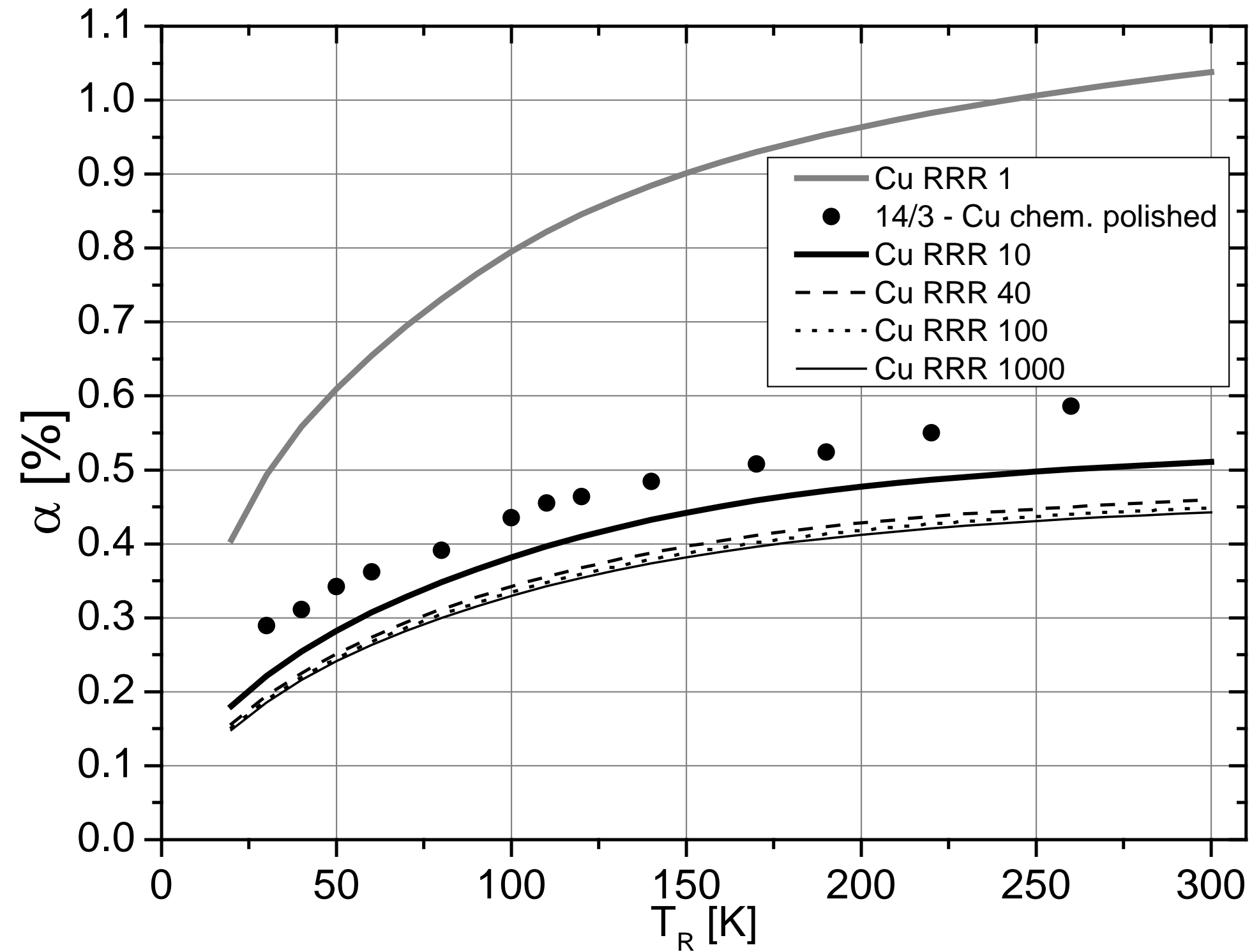

Fig. 2 Lines - absorptivity of copper with various RRR, points - measured absorptivity of chemically polished Cu with RRR = 40 for bulk material.

4.2.2. Accidental impurities

Fig. 3 shows absorptivity of the copper coated with Ni or Au as a protective layer. We can see multiple increase of absorptivity after coating of the copper. Analysis revealed 10 % of phosphorus and 0.6 % of cobalt in commercial chemically deposited Ni and galvanical Au layer, respectively. The presence and the effect of such alloying elements may not be evident when the decision about the deposition technology is done.

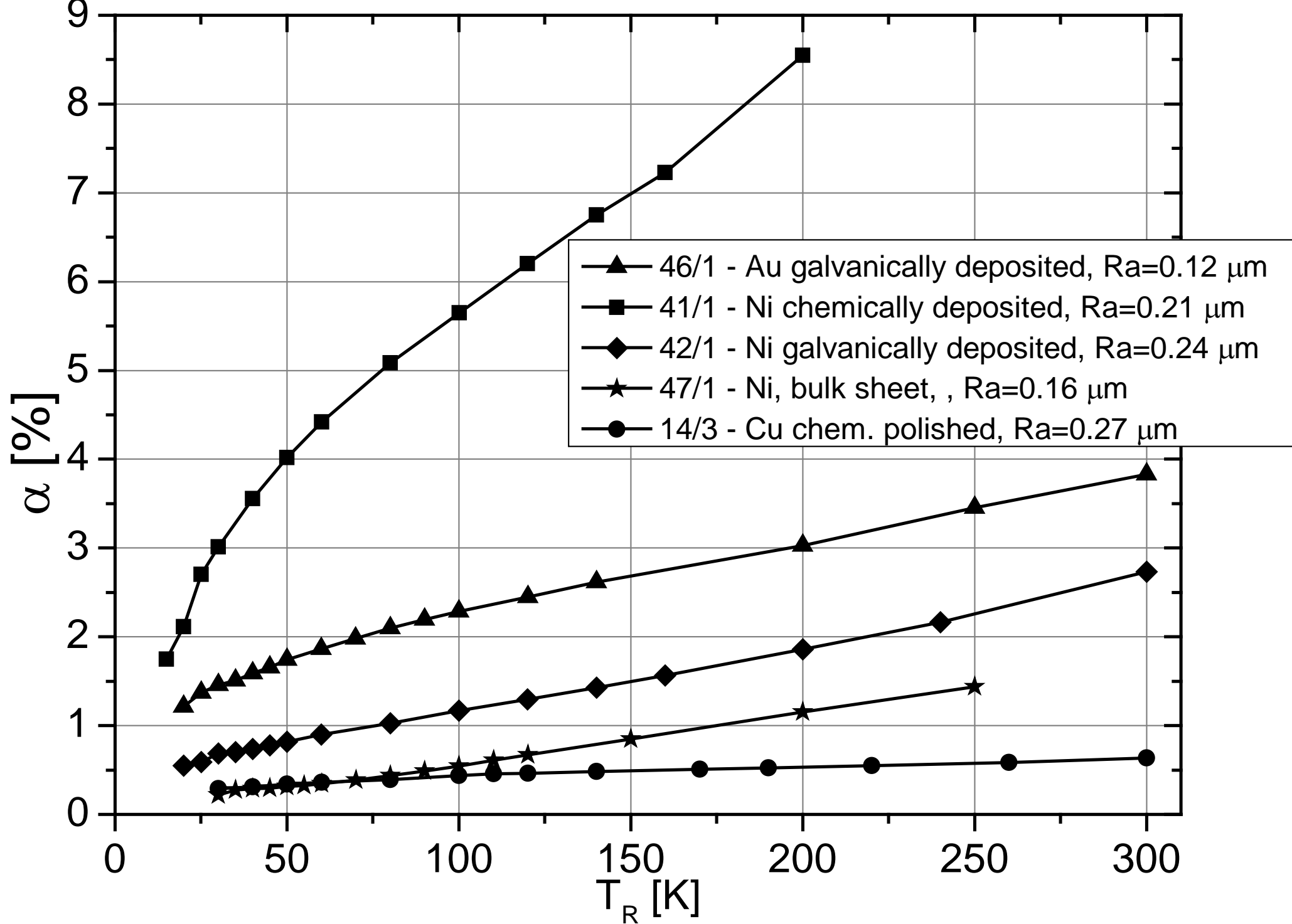

Fig. 3 Absorptivity of galvanically deposited layers of Ni and Au are compared with chemically deposited Ni, bulk Ni sheet and chemically polished Cu sheet.

### 4.3. Chemical treatment

The lowest values of absorptivity and emissivity of aluminium and copper we have achieved by chemical polishing with a mixture of concentrated acids (Table 3, bath 1). The chemical polishing bath may not be applicable for larger objects. Etching by diluted nitric acid could be used to remove damaged or contaminated surface layer. Nevertheless this treatment gives slightly higher absorptivity (Fig. 4) in comparison with chemical polishing. Such behaviour may be related to selectivity of the etching process.
The chemically polished copper surface oxidizes quickly on the air and it is prone to be stained by fingerprints. As a protection of such a surface a controlled oxidation (passivation) can be done (Table 3, bath 3). The impact of the passivation of Cu in comparison with the chemically polished Cu is shown in Fig. 4.

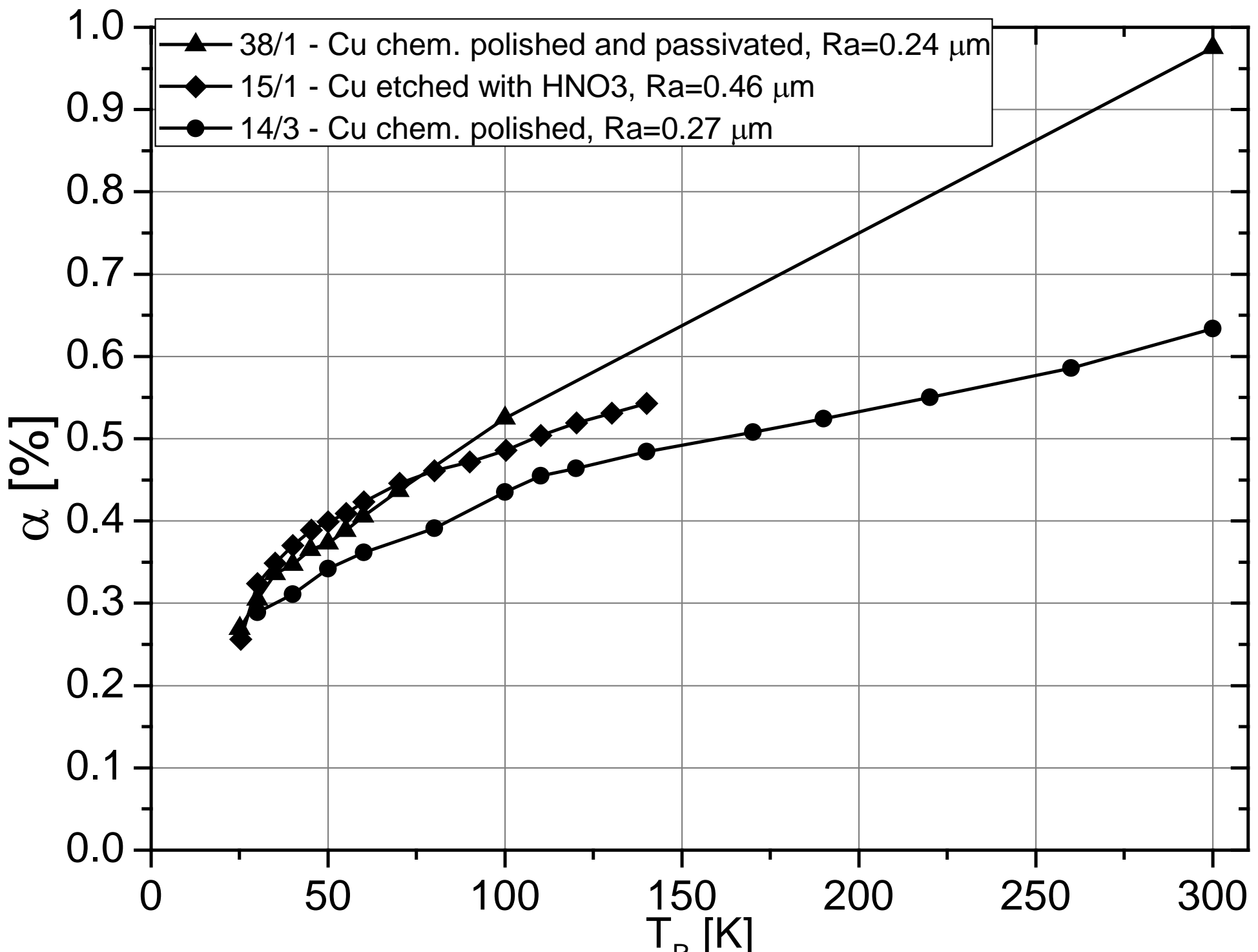

Fig. 4 Absorptivity of etched and passivated Cu is compared with absorptivity of chemically polished Cu.

### 4.4. Mechanical working

Many surfaces radiating or absorbing thermal radiation are finished by mechanical machining. Fig. 5 illustrates the impact of turning, abrasion, and mechanical polishing on the absorptivity of Cu and Al. The increase of the absorptivity after mechanical treatment is caused by mechanical damaging of the surface layer. This statement is supported by measurement of the turned sample after annealing at 650°C for 30 minutes which reduced the absorptivity markedly. The measurements also clearly show that mechanical polishing to high gloss, especially when used as a last step of surface treatment, is an unreasonable way to reduce the absorptivity.

### 4.5. Surface contamination

Sometimes it is not possible to use chemical or electrochemical polishing process to reduce its absorptivity or emissivity. The plot in Fig. 6 shows absorptivity of untreated surfaces of aluminium and copper. The samples made of rolled sheets were only washed with acetone to remove grease, oil and dust. The absorptivity of untreated surface is more than three times higher than the absorptivity of chemically polished samples but it is only by ~20 - 25 % higher than the absorptivity of mechanically polished Cu and Al (Fig. 5). Many cryogenic systems contain parts made of plastics kept at room temperature. These materials (epoxies, glues, insulations, composites) have a capability to absorb water and release it slowly in vacuum. This water precipitates on cold parts as a thin layer of ice. The impact of such layer (Hanzelka et al., 2010) on the absorptivity of chemically polished Al is also shown in the Fig. 6. A layer of 85 nm ice has doubled the absorptivity of the chemically polished Al. The absorptivity of Al with this ice layer is practically the same as the absorptivity of the untreated surface.

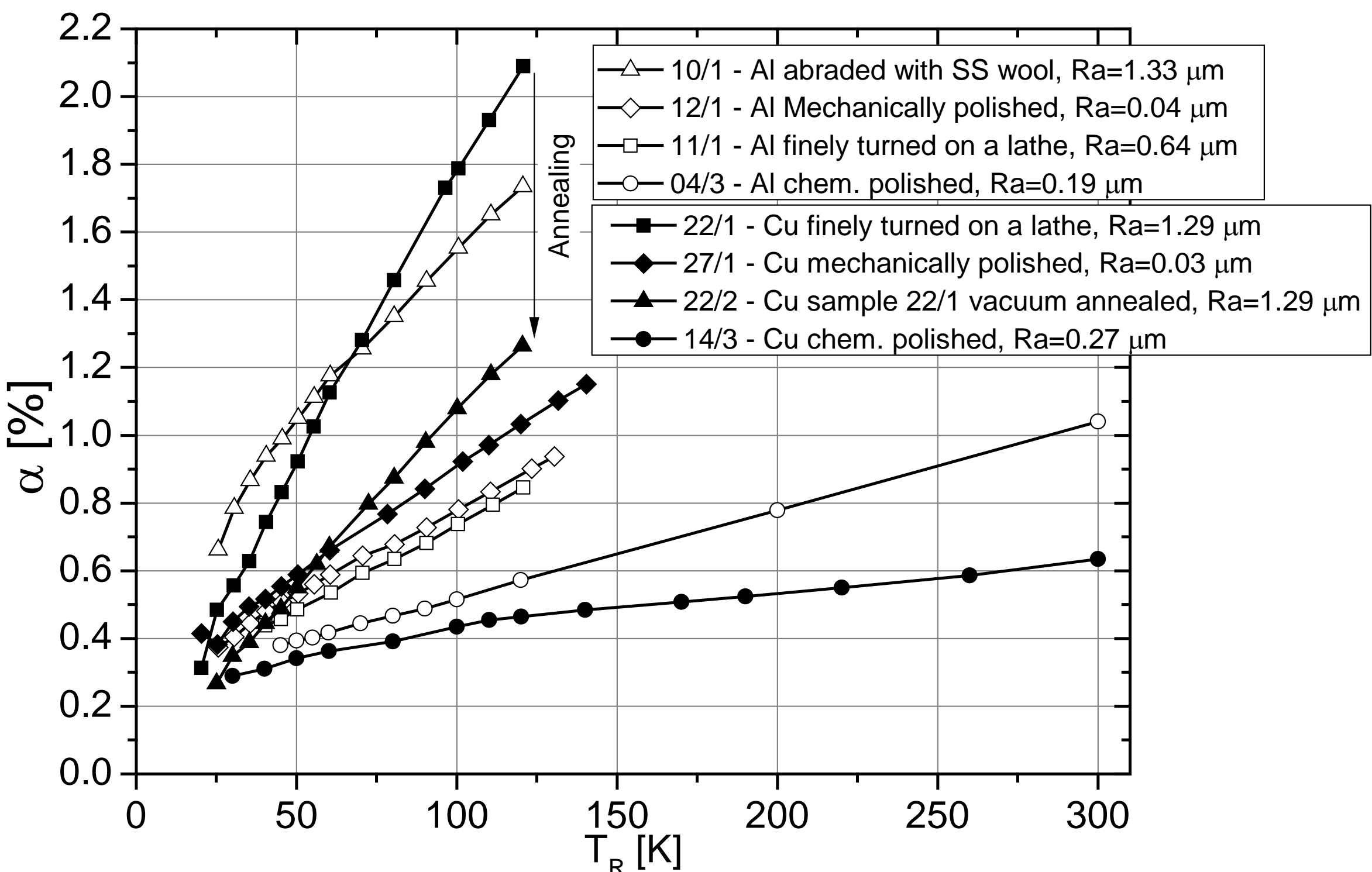

Fig. 5 Effect of surface machining on absorptivity of Al and Cu. The impact of the working of Cu is reduced by annealing.

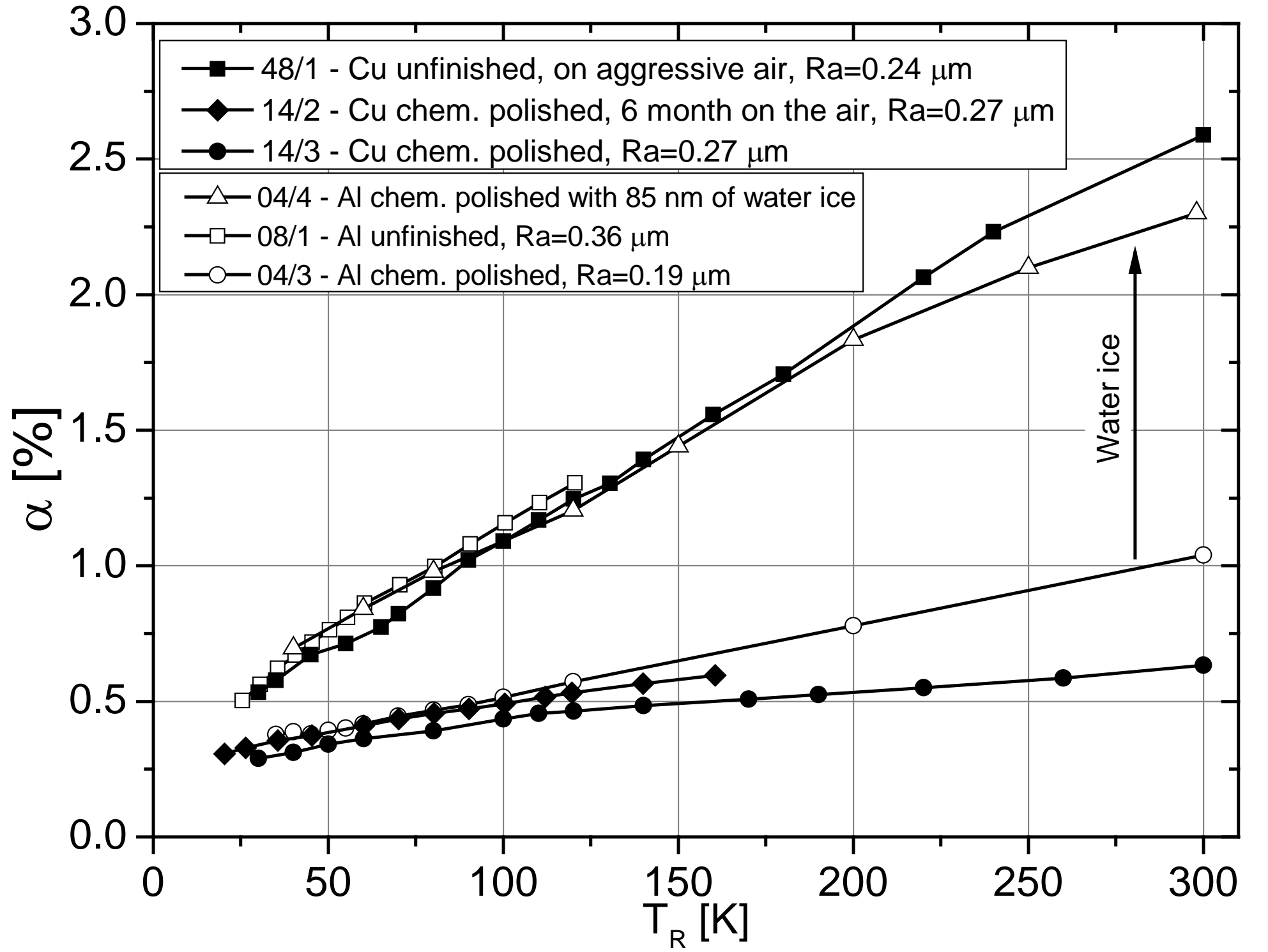

Fig. 6 Influence of the natural oxide layer and ice layer on the absorptivity compared with chemically polished surface

### 4.6. Roughness

The influence of surface roughness on the thermal radiative properties is one of the discussed topics when the choice of the technology of the surface treatment has to be done. The range of wavelengths of low temperature thermal radiation from ~ 10 μm to 300 μm hints that surfaces need not to be visually smooth (mirror like). Unfortunately here any consideration about thermal radiative properties from the visual appearance fails.

The roughness Ra of the samples presented in this contribution, measured by industrial apparatus MarSurf [MARSURF], is tabulated in Table 2. The samples in Table 2 can be characterised as nearly optically smooth (Ra to wavelength ratio << 1).

Small influence of roughness of chemically treated surfaces (Ra < 0.5) is obvious from comparison of absorptivity values with theoretical absorptivity of smooth surface of Cu (Domoto et al., 1970) or from agreement between the measured absorptivity of chemically polished Al and that calculated from optical constants (Rakic, 1995), e.g.

## 5. CONCLUSIONS

Thermal radiative properties of metals generally depend on the temperature of the material and on the temperature of the source of thermal radiation (Fig. 1). When searching for thermal radiative properties of metals in the literature the information about the temperature should be taken with care. Reduction of absorptivity by using an extra pure material is not effective way to reduce the absorption of the thermal radiation (Fig. 2). On the other hand small amount of accidental impurities in the surface metallic coating can increase the absorptivity (Fig. 3). As only the very thin surface layer of metals interacts with thermal radiation it suggests itself to remove an impaired surface layer chemically. A difference between etching and chemical polishing was observed (Fig. 4).

Protective passivation of copper done by controlled oxidation in the bath 3 does not increase the absorptivity substantially. Surprisingly naturally contaminated surfaces (Fig. 6) of the rolled sheets proved lower absorptivity than the mechanically worked surfaces (Fig. 5). Thin layer of water ice having source in the warm parts of the cryostat and deposited on cold parts can have the same or stronger influence on the absorptivity the natural oxide layer (Fig. 6).

Damage of the metal structure caused by mechanical stress applied on the surface by cutting tools or by abrasion has significant impact on the absorptivity. The absorptivity could be reduced by annealing of the machined parts if possible (turned Cu in Fig. 5).

Measurements of the absorption of the thermal radiation at low temperatures do not suggest marked effect of surface roughness (Ra < 1 μm) on absorptivity.

## 6. REFERENCES

DOMOTO, G. A., BOEHM, R. F. & TIEN, C. L. 1970. Experimental Investigation of Radiative Transfer between Metallic Surfaces at Cryogenic Temperatures. Journal of Heat Transfer, 92, 412-&.

HANZELKA, P., MUSILOVA, V. & KRALIK, T. 2010. Influence of condensed water on heat radiation absorptivity at cryogenic temperatures. Cryogenics, 50, 331-335.

KRALIK, T. H., P.; MUSILOVA, V.; SRNKA, A. 2004. Device for measurement of thermal emissivity at cryogenic temperatures. Proceedings of 8th Cryogenics conference 2004, 2004 Prague. Icaris.

MUSILOVA, V., HANZELKA, P., KRALIK, T. & SRNKA, A. 2005. Low temperature radiative properties of materials used in cryogenics. Cryogenics, 45, 529-536.

MUSILOVA, V., KRALIK, T., HANZELKA, P. & SRNKA, A. 2007. Effect of different treatments of copper surface on its total hemispherical absorptivity bellow 77 K. Cryogenics, 47, 257-261.

RAKIC, A. D. 1995. Algorithm for the determination of intrinsic optical constants of metal films: application to aluminum. Applied Optics, 34, 4755-4767.

RAMANATHAN, K. G. 1952. Infra-Red Absorption by Metals at Low Temperatures. Proceedings of the Physical Society. Section A, 65, 532.

SIEGEL, R. & HOWELL, J. R. 2002. Thermal radiation heat transfer, New York, Taylor & Francis.

MARSURF, Industrial apparatus for roughness measurement, Evaluation unit: MarSurf M300, Drive unit: MarSurf RD18, Probe: PHT 6-350 with probe tip geometry 2 μm/90° (diamond)

# 7. ACKNOWLEDGEMENT

The authors acknowledge the support from Ministry of Education, Youth and Sports of the Czech Republic (LO1212) together with the European Commission (ALISI No. CZ.1.05/2.1.00/01.0017) and Grant Agency of the Czech Republic (GA CR) project No. 14-07397S.